\documentclass[a4paper,11pt]{article}
\usepackage{pos}

\usepackage{textcomp}
\usepackage{amsfonts} 
\usepackage{amsmath} 
\usepackage{slashed}
\usepackage{pifont}
\usepackage{multirow,lscape}
\usepackage{array}
\usepackage{subcaption}
\usepackage{verbatim}
\usepackage{bm}
\usepackage{comment}
\usepackage{xcolor}
\usepackage{url} 

\graphicspath{{./fig/}}

\providecommand{\wbar}[1]{\overline#1}

\providecommand{\mate}[3]{\langle#1\lvert#2\rvert#3\rangle}

\renewcommand{\Re}{\mathrm{Re}\,}
\renewcommand{\Im}{\mathrm{Im}\,}

\providecommand{\GeV}{\;\mathrm{GeV}}

\definecolor{HLBlue}{HTML}{6599FF}
\definecolor{HLOrange}{HTML}{FF6600}

\newcommand{\BK}{\hat{B}_{K}}
\newcommand{\Vcb}{|V_{cb}|}

\newcommand{\Vus}{|V_{us}|}

\newcommand{\eps}{\varepsilon}

\newcommand{\epsK}{\varepsilon_{K}}

\newcommand{\BtoDstp}{\bar{B} \to D^{(\ast)} \ell \bar{\nu}}
\newcommand{\BtoDst}{\bar{B} \to D^\ast \ell \bar{\nu}}

\newcommand{\red}[1]{\textcolor{red}{#1}} 
\newcommand{\blue}[1]{\textcolor{blue}{#1}}

\begin{document}
\title{2022 Update on $\epsK$ with lattice QCD inputs}
\ShortTitle{$\epsK$ with lattice QCD inputs}

\author[a,1]{Sunghee Kim}
\author[a,1]{Sunkyu Lee}
\author*[a,b,1]{Weonjong Lee}
\author[b,c,1]{Jaehoon Leem}
\author[d,1]{Sungwoo Park}

\affiliation[a]{Lattice Gauge Theory Research Center, CTP, and FPRD,
  Department of Physics and Astronomy, \\
  Seoul National University,
  Seoul 08826, South Korea}


\affiliation[b]{School of Physics,
  Korea Institute for Advanced Study (KIAS),
  Seoul 02455, South Korea}

\affiliation[c]{Computational Science and Engineering Team,
  Innovation Center, Samsung Electronics, Hwaseong,
  Gyeonggi-do 18448, South Korea.}

\affiliation[d]{
  Thomas Jefferson National Accelerator Facility,
  12000 Jefferson Avenue,
  Newport News, VA 23606, USA}

\emailAdd{wlee@snu.ac.kr}

\note{The SWME collaboration}

\abstract{
  We present recent updates for $\epsK$ determined directly from the
  standard model (SM) with lattice QCD inputs such as $\BK$, $\Vcb$,
  $\Vus$, $\xi_0$, $\xi_2$, $\xi_\text{LD}$, $f_K$, and $m_c$.  We
  find that the standard model with exclusive $\Vcb$ and other lattice
  QCD inputs describes only 65\% of the experimental value of
  $|\epsK|$ and does not explain its remaining 35\%, which leads to a
  strong tension in $|\epsK|$ at the $5.1\sigma \sim 3.9\sigma$ level
  between the SM theory and experiment.  We also find that this
  tension disappears when we use the inclusive value of $\Vcb$
  obtained using the heavy quark expansion based on the QCD sum rule
  approach, although this inclusive tension is small ($\approx
  1.4\sigma$) but keeps increasing as time goes on. }

\FullConference{%
The 39th International Symposium on Lattice Field Theory,\\
8th-13th August, 2022,\\
Rheinische Friedrich-Wilhelms-Universität Bonn, Bonn, Germany
}

\maketitle

\section{Introduction}
This paper is an update from our previous reports \cite{ Lee:2021crz,
  Kim:2019vic, Bailey:2018feb, Bailey:2015tba, Bailey:2018aks,
  Jang:2017ieg, Bailey:2015frw}.
Here, we present recent progress in determination of $|\epsK|$ with
updated inputs from lattice QCD.
Updated input parameters include $\bar{\rho}$, $\bar{\eta}$, exclusive
$\Vcb$, inclusive $\Vcb$, $M_W$, and $M_t$.

Here, we follow the color convention of our previous papers \cite{
  Lee:2021crz, Kim:2019vic, Bailey:2018feb, Bailey:2015tba,
  Bailey:2018aks, Jang:2017ieg, Bailey:2015frw} in Tables
\ref{tab:input-WP-eta}--\ref{tab:epsK}.
We use the red color for the new input data which is used to evaluate
$\epsK$.
We use the blue color for the new input data which is not used for
some obvious reason.

\section{Input parameters: Wolfenstein parameters}
\label{sec:wp}
In Table \ref{tab:input-WP-eta}\,(\subref{tab:WP}), we present the
most updated Wolfenstein parameters available in the market.
As explained in Ref.~\cite{ Bailey:2018feb, Bailey:2015frw}, we use
the results of angle-only-fit (AOF) in Table
\ref{tab:input-WP-eta}\,(\subref{tab:WP}) in order to avoid unwanted
correlation between $(\epsK, \Vcb)$, and $(\bar\rho, \bar\eta)$.
We determine $\lambda$ from $\Vus$ which is obtained from the $K_{\ell
  2}$ and $K_{\ell 3}$ decays using lattice QCD inputs for form
factors and decay constants as explained in Ref.~\cite{
  FlavourLatticeAveragingGroupFLAG:2021npn}.
We determine the $A$ parameter from $\Vcb$.
\begin{table}[h!]
  \begin{subtable}{0.73\linewidth}
    \renewcommand{\arraystretch}{1.2}
    \resizebox{1.0\linewidth}{!}{
      \begin{tabular}{ @{\qquad} c @{\qquad} | l l | l l | l l }
        \hline\hline
        WP
        & \multicolumn{2}{c|}{CKMfitter}
        & \multicolumn{2}{c|}{UTfit}
        & \multicolumn{2}{c}{AOF}
        \\ \hline
        $\lambda$
        & $0.22475(25)$       & \cite{Charles:2004jd}
        & $0.22500(100)$      & \cite{Bona:2006ah}
        & \red{ $0.2249(5)$ } & \cite{FlavourLatticeAveragingGroupFLAG:2021npn}
        \\ \hline
        $\bar{\rho}$
        & $0.1577(96)$        & \cite{Charles:2004jd}
        & $0.148(13)$         & \cite{Bona:2006ah}
        & \red{ $0.156(17)$ } & \cite{UTfit:2022hsi}
        \\ \hline
        $\bar{\eta}$
        & $0.3493(95)$        & \cite{Charles:2004jd}
        & $0.348(10)$         & \cite{Bona:2006ah}
        & \red{ $0.334(12)$ } & \cite{UTfit:2022hsi}
        \\ \hline\hline
      \end{tabular}
    } 
    \caption{Wolfenstein parameters}
    \label{tab:WP}
  \end{subtable} 
  \hfill
  \begin{subtable}{0.26\linewidth}
    \renewcommand{\arraystretch}{1.3}
    \resizebox{1.0\linewidth}{!}{
      \begin{tabular}[b]{ c l c }
        \hline\hline
        Input & Value & Ref.
        \\ \hline
        $\eta_{cc}$ & $1.72(27)$   & \cite{Bailey:2015tba}
        \\
        $\eta_{tt}$ & $0.5765(65)$ & \cite{Buras2008:PhysRevD.78.033005}
        \\
        $\eta_{ct}$ & $0.496(47)$  & \cite{Brod2010:prd.82.094026}
        \\ \hline\hline
      \end{tabular}
    } 
    \caption{$\eta_{ij}$}
    \label{tab:eta}
  \end{subtable} 
  \caption{ (\subref{tab:WP}) Wolfenstein parameters and
    (\subref{tab:eta}) QCD corrections: $\eta_{ij}$ with $i,j = c,t$.}
  \label{tab:input-WP-eta}
\end{table}

\section{Input parameters: $\Vcb$}
\label{sec:Vcb}
In Table \ref{tab:Vcb} (\subref{tab:ex-Vcb}) and
(\subref{tab:in-Vcb}), we present recently updated results for
exclusive $\Vcb$ and inclusive $\Vcb$ respectively.
In Table \ref{tab:Vcb} (\subref{tab:ex-Vcb}), we summarize results for
exclusive $\Vcb$ obtained by various groups: HFLAV, BELLE, BABAR,
FNAL/MILC, LHCb, and FLAG.
Results from LHCb comes from analysis on $B_s\to D^*_s \ell \bar{\nu}$
decays which are not available in the $B$-factories.
Since results for $B_s$ decay channels have poor statistics, we drop
out them here without loss of fairness.
The rest of results for exclusive $\Vcb$ have comparable size of
errors and are consistent with one another within $1.0\sigma$.
In addition, we find that the results are consistent between
the CLN and BGL analysis, after the clamorous debates \cite{
  Bailey:2018feb, FermilabLattice:2021cdg}.

In Table \ref{tab:Vcb} (\subref{tab:in-Vcb}), we present recent
results for inclusive $\Vcb$.
The Gambino group has reported updated results for inclusive $\Vcb$
in 2021.
There are a number of attempts to calculate inclusive $\Vcb$ in
lattice QCD, but they belong to a category of exploratory study rather
than that of precision measurement yet \cite{ Barone:2022gkn}.
\begin{table}[t!]
  \begin{subtable}{1.0\linewidth}
    \renewcommand{\arraystretch}{1.2}
    \center
    \resizebox{1.0\textwidth}{!}{
      \begin{tabular}{@{\qquad} l @{\qquad} l @{\qquad} l @{\qquad} l @{\qquad} l @{\qquad}}
        \hline\hline
        channel & value & method & ref & source \\ \hline
        ex-comb & \red{$39.25(56)$} & CLN & \cite{HFLAV:2019otj} p115e223 & HFLAV-2021  
        \\ \hline
        $B\to D^* \ell \bar{\nu}$
        & \red{$39.0(2)(6)(6)$} & CLN & \cite{Belle:2018ezy} erratum p4 & BELLE-2021 \\
        $B\to D^* \ell \bar{\nu}$
        & \red{$38.9(3)(7)(6)$} & BGL & \cite{Belle:2018ezy} erratum p4 & BELLE 2021
        \\ \hline
        $B\to D^* \ell \bar{\nu}$
        & \red{$38.40(84)$} & CLN & \cite{ BaBar:2019vpl} p5t2 & BABAR-2019
        \\
        $B\to D^* \ell \bar{\nu}$
        & \red{$38.36(90)$} & BGL & \cite{ BaBar:2019vpl} p5t1 & BABAR-2019
        \\ \hline
        $B\to D^* \ell \bar{\nu}$
        & \red{$38.40(78)$} & BGL & \cite{ FermilabLattice:2021cdg} p27e76 &
        FNAL/MILC-2022
        \\ \hline
        $B_s\to D^*_s \ell \bar{\nu}$
        & \blue{$41.4(6)(9)(12)$} & CLN & \cite{ LHCb:2020cyw} p15 & LHCb-2020
        \\
        $B_s\to D^*_s \ell \bar{\nu}$
        & \blue{$42.3(8)(9)(12)$} & BGL & \cite{ LHCb:2020cyw} p15 & LHCb-2020
        \\ \hline
        ex-comb
        & \textcolor{red}{$39.48(68)$} & comb & \cite{ FlavourLatticeAveragingGroupFLAG:2021npn} p145 & FLAG-2021
        \\ \hline\hline
      \end{tabular}
    } 
    \caption{Exclusive $\Vcb$ in units of $10^{-3}$.}
    \label{tab:ex-Vcb}
  \end{subtable}
  \begin{subtable}{1.0\linewidth}
    \renewcommand{\arraystretch}{1.2}
    \center
    \vspace*{+2mm}
    \resizebox{1.0\textwidth}{!}{
      \begin{tabular}{ @{\qquad} l @{\qquad\qquad} l @{\qquad\qquad\qquad} l @{\qquad\qquad} l @{\qquad} }
        \hline\hline
        channel        & value         & ref  & source \\ \hline
        kinetic scheme & \blue{$42.16(51)$} & \cite{ Bordone:2021oof} p1 & Gambino-2021
        \\
        kinetic scheme & \blue{$42.00(64)$} & \cite{ FlavourLatticeAveragingGroupFLAG:2021npn, Gambino:2016jkc} p145 & FLAG-2021 
        \\ \hline
        1S scheme      & \red{$41.98(45)$} & \cite{ HFLAV:2019otj} p110e208 & HFLAV-2021 
        \\ \hline\hline
      \end{tabular}
    } 
    \caption{Inclusive $\Vcb$ in units of $10^{-3}$.}
    \label{tab:in-Vcb}
  \end{subtable}
  \caption{ Results for (\subref{tab:ex-Vcb}) exclusive $\Vcb$ and
    (\subref{tab:in-Vcb}) inclusive $\Vcb$. The p115e223 is an
    abbreviation for Eq.~(223) in page 115. The p5t2 is an
    abbreviation for Table 2 in page 5.}
  \label{tab:Vcb}
\end{table}
\section{Input parameter $\xi_0$}
The absorptive part of long distance effects on $\epsK$ is parametrized
into $\xi_0$.
\begin{align}
  \xi_0  &= \frac{\Im A_0}{\Re A_0}, \qquad
  \xi_2 = \frac{\Im A_2}{\Re A_2}, \qquad
  \Re \left(\frac{\eps'}{\eps} \right) =
  \frac{\omega}{\sqrt{2} |\eps_K|} (\xi_2 - \xi_0) \,.
  \label{eq:e'/e:xi0}
\end{align}
There are two independent methods to determine $\xi_0$ in lattice QCD:
the indirect and direct methods.
The indirect method is to determine $\xi_0$ using
Eq.~\eqref{eq:e'/e:xi0} with lattice QCD results for $\xi_2$ combined
with experimental results for $\eps'/\eps$, $\epsK$, and $\omega$.
The direct method is to determine $\xi_0$ directly using the lattice
QCD results for $\Im A_0$, combined with experimental results for $\Re
A_0$.

In Table~\ref{tab:xi0-sum} (\subref{tab:exp-ReA0-ReA2-1}), we summarize
experimental results for $\Re A_0$ and $\Re A_2$.
In Table~\ref{tab:xi0-sum} (\subref{tab:ImA0-ImA2-1}), we summarize
lattice results for $\Im A_0$ and $\Im A_2$ calculated by RBC-UKQCD.
In Table~\ref{tab:xi0-sum} (\subref{tab:xi0-1}), we summarize results
for $\xi_0$ which is obtained using results in Table~\ref{tab:xi0-sum}
(\subref{tab:exp-ReA0-ReA2-1}) and (\subref{tab:ImA0-ImA2-1}).

Here, we use results of the indirect method for $\xi_0$ to evaluate
$\epsK$, since its systematic and statistical errors are much smaller
than those of the direct method.

\begin{table}[htbp]
  \begin{subtable}{1.0\linewidth}
    \renewcommand{\arraystretch}{1.2}
    \center
    \resizebox{1.0\textwidth}{!}{
      \begin{tabular}{ @{\qquad} l @{\qquad} l @{\qquad\qquad} l @{\qquad\qquad} l @{\qquad} l @{\qquad\qquad} }
        \hline\hline
        parameter & method & value & Ref. & source \\ \hline
        $\Re A_0$ & exp & \red{ $3.3201(18) \times 10^{-7} \GeV$ } &
        \cite{ Blum:2015ywa, Bai:2015nea}  & NA
        \\
        $\Re A_2$ & exp & \red{ $1.4787(31) \times 10^{-8} \GeV$ } &
        \cite{ Blum:2015ywa} & NA
        \\ \hline
        $\omega$ & exp & $0.04454(12)$ &
        \cite{ Blum:2015ywa} & NA
        \\ \hline
        $|\epsK|$ & exp & $2.228(11) \times 10^{-3}$ &
        \cite{ Zyla:2020zbs} & PDG-2021
        \\
        $\Re(\eps'/\eps)$ & exp & $1.66(23) \times 10^{-3}$ &
        \cite{ Zyla:2020zbs} & PDG-2021
        \\ \hline\hline
      \end{tabular}
    } 
    \caption{Experimental results for $\omega$, $\Re A_0$ and $\Re A_2$.}
    \label{tab:exp-ReA0-ReA2-1}
  \end{subtable}
  \vspace*{4mm}
  \begin{subtable}{1.0\linewidth}
    \renewcommand{\arraystretch}{1.2}
    \center
    \resizebox{1.0\textwidth}{!}{
      \begin{tabular}{ @{\qquad} l @{\qquad} l @{\qquad}@{\qquad} l @{\qquad}@{\qquad} l @{\qquad} l }
        \hline\hline
        parameter & method & value ($\GeV$) & Ref. & source \\ \hline
        $\Im A_0$ & lattice & \red{$-6.98(62)(144) \times 10^{-11}$} &
        \cite{ RBC:2020kdj} p4t1   & RBC-UK-2020 
        \\
        $\Im A_2$ & lattice & \red{$-8.34(103) \times 10^{-13}$}  &
        \cite{ RBC:2020kdj} p31e90 & RBC-UK-2020 
        \\ \hline\hline
      \end{tabular}
    } 
    \caption{Results for $\Im A_0$, and $\Im A_2$ in lattice QCD. }
    \label{tab:ImA0-ImA2-1}
  \end{subtable}
  \vspace*{3mm}
  \begin{subtable}{1.0\linewidth}
    \renewcommand{\arraystretch}{1.2}
    \center
    \resizebox{1.0\textwidth}{!}{
      \begin{tabular}{@{\qquad} l @{\qquad\qquad} l @{\qquad\qquad} l @{\qquad\qquad} l @{\qquad\qquad} l @{\qquad} }
        \hline\hline
        parameter & method & value & ref & source \\ \hline
        $\xi_0$ & indirect & $-1.738(177) \times 10^{-4}$ & \cite{ RBC:2020kdj} & SWME \\
        $\xi_0$ & direct  & $-2.102(472) \times 10^{-4}$  & \cite{ RBC:2020kdj} & SWME \\ \hline\hline
      \end{tabular}
    } 
    \caption{Results for $\xi_0$ obtained using the direct and indirect
      methods in lattice QCD. }
    \label{tab:xi0-1}
  \end{subtable}
  \caption{Results for $\xi_0$. Here, we use the same notation as in
    Table \ref{tab:Vcb}.}
  \label{tab:xi0-sum}
\end{table}

\section{Input parameters: $\BK$, $\xi_\text{LD}$, and others}
In FLAG 2021 \cite{ FlavourLatticeAveragingGroupFLAG:2021npn}, they
report lattice QCD results for $\BK$ with $N_f=2$, $N_f=2+1$, and $N_f
= 2+1+1$.
Here, we use the results for $\BK$ with $N_f=2+1$, which is obtained
by taking an average over the four data points from BMW 11, Laiho 11,
RBC-UKQCD 14, and SWME 15 in Table
\ref{tab:input-BK-other}\;(\subref{tab:BK}).

\begin{table}[htbp]
  \begin{subtable}{0.40\linewidth}
    \renewcommand{\arraystretch}{1.45}
    \resizebox{1.0\linewidth}{!}{
      \begin{tabular}{ l  c  l }
        \hline\hline
        Collaboration & Ref. & $\BK$  \\ \hline
        SWME 15       & \cite{Jang:2015sla} & $0.735(5)(36)$     \\
        RBC/UKQCD 14  & \cite{Blum:2014tka} & $0.7499(24)(150)$  \\
        Laiho 11      & \cite{Laiho:2011np} & $0.7628(38)(205)$  \\
        BMW 11        & \cite{Durr:2011ap}  & $0.7727(81)(84)$  \\ \hline
        FLAG 2021     & \cite{FlavourLatticeAveragingGroupFLAG:2021npn}
                                            & $0.7625(97)$
        \\ \hline\hline
      \end{tabular}
    } 
    \caption{$\BK$}
    \label{tab:BK}
  \end{subtable} 
  \hfill
  \begin{subtable}{0.60\linewidth}
    \renewcommand{\arraystretch}{1.2}
    \resizebox{1.0\linewidth}{!}{
      \begin{tabular}{ @{\qquad} c @{\qquad} l @{\qquad} l @{\qquad} }
        \hline\hline
        Input & Value & Ref. \\ \hline
        $G_{F}$
        & $1.1663787(6) \times 10^{-5}$ GeV$^{-2}$
        & PDG-22 \cite{ Workman:2022ynf} \\ \hline
        $M_{W}$
        & \red{ $80.356(6)$ GeV }
        & SM-22 \cite{ Workman:2022ynf}\\ \hline
        $\theta$
        & $43.52(5)^{\circ}$
        & PDG-22 \cite{ Workman:2022ynf} \\ \hline
        $m_{K^{0}}$
        & $497.611(13)$ MeV
        & PDG-22 \cite{ Workman:2022ynf} \\ \hline
        $\Delta M_{K}$
        & $3.484(6) \times 10^{-12}$ MeV
        & PDG-22 \cite{ Workman:2022ynf} \\ \hline
        $F_K$
        & $155.7(3)$ MeV
        & FLAG-21 \cite{ FlavourLatticeAveragingGroupFLAG:2021npn}
        \\ \hline\hline
      \end{tabular}
    } 
    \caption{Other parameters}
    \label{tab:other}
  \end{subtable} 
  \caption{ (\subref{tab:BK}) Results for $\BK$ and
    (\subref{tab:other}) other input parameters.}
  \label{tab:input-BK-other}
\end{table}

The dispersive long distance (LD) effect is defined as
\begin{align}
  \xi_\text{LD} &=  \frac{m^\prime_\text{LD}}{\sqrt{2} \Delta M_K} \,,
  \qquad
  m^\prime_\text{LD}
  = -\Im \left[ \mathcal{P}\sum_{C}
    \frac{\mate{\wbar{K}^0}{H_\text{w}}{C} \mate{C}{H_\text{w}}{K^0}}
         {m_{K^0}-E_{C}}  \right]
  \label{eq:xi-LD}
\end{align}
As explained in Refs.~\cite{ Bailey:2018feb}, there are two
independent methods to estimate $\xi_\text{LD}$: one is the BGI
estimate \cite{ Buras:2010}, and the other is the RBC-UKQCD estimate
\cite{ Christ:2012, Christ:2014qwa}.
The BGI method is to estimate the size of $\xi_\text{LD}$ using
chiral perturbation theory as follows,
\begin{align}
  \xi_\text{LD} &= -0.4(3) \times \frac{\xi_0}{ \sqrt{2} }
  \label{eq:xiLD:bgi}
\end{align}
The RBC-UKQCD method is to estimate the size of $\xi_\text{LD}$
as follows,
\begin{align}
  \xi_\text{LD} &= (0 \pm 1.6)\%.
  \label{eq:xiLD:rbc}
\end{align}
Here, we use both methods to estimate the size of $\xi_\text{LD}$.

In Table \ref{tab:input-WP-eta}\;(\subref{tab:eta}), we present higher
order QCD corrections: $\eta_{ij}$ with $i,j = t,c$.
A new approach using $u-t$ unitarity instead of $c-t$ unitarity
appeared in Ref.~\cite{ Brod:2019rzc}, which is supposed to have a
better convergence with respect to the charm quark mass.
But we have not incorporated this into our analysis yet, which we will
do in near future.

In Table \ref{tab:input-BK-other}\;(\subref{tab:other}), we present other
input parameters needed to evaluate $\epsK$.

\section{Quark masses}
In Table \ref{tab:m_c:m_t}, we present the charm quark mass $m_c(m_c)$
and top quark mass $m_t(m_t)$.
From FLAG 2021 \cite{ FlavourLatticeAveragingGroupFLAG:2021npn}, we
take the results for $m_c (m_c)$ with $N_f = 2+1$, since there is some
inconsistency among the lattice results of various groups with $N_f =
2+1+1$.
For the top quark mass, we use the PDG 2022 results for the pole mass
$M_t$ to obtain $m_t (m_t)$.
\begin{table}[t!]
  \begin{subtable}{0.46\linewidth}
    \vspace*{-5mm}
    \renewcommand{\arraystretch}{1.4}
    \resizebox{0.99\linewidth}{!}{
      \begin{tabular}{ l l l l }
        \hline\hline
        {\small Collaboration} & $N_f$ & $m_c(m_c)$ & Ref.
        \\ \hline
        FLAG 2021       & $2+1$   & \red{$1.275(5)$}
        & \cite{FlavourLatticeAveragingGroupFLAG:2021npn}
        \\
        FLAG 2021       & $2+1+1$ & $1.278(13)$
        & \cite{FlavourLatticeAveragingGroupFLAG:2021npn}
        \\ \hline\hline
      \end{tabular}
    } 
    \caption{$m_c(m_c)$ [GeV]}
    \label{tab:m_c}
  \end{subtable} 
  \hfill
  \begin{subtable}{0.52\linewidth}
    \renewcommand{\arraystretch}{1.2}
    \vspace*{-5mm}
    \resizebox{0.99\linewidth}{!}{
      \begin{tabular}{ l l l l }
        \hline\hline
        {\small Collaboration} & $M_t$ & $m_t(m_t)$ & Ref.
        \\ \hline
        PDG 2019 & 172.9(4)   & 163.08(38)(17)         & \cite{Tanabashi:2018oca} \\
        PDG 2021 & 172.76(30) & 162.96(28)(17)         & \cite{Zyla:2020zbs}
        \\
        PDG 2022 & 172.69(30) & \red{ 162.90(28)(17) } & \cite{Workman:2022ynf}
        \\ \hline\hline
      \end{tabular}
    } 
    \caption{ $m_t(m_t)$ [GeV] }
    \label{tab:m_t}
  \end{subtable} 
  \caption{  Results for (\subref{tab:m_c}) charm quark mass and
    (\subref{tab:m_t}) top quark mass. }
  \label{tab:m_c:m_t}
\end{table}

\begin{table}[h!]
  \begin{subfigure}{0.55\linewidth}
    \vspace{-3mm}
    \includegraphics[width=1.0\linewidth]{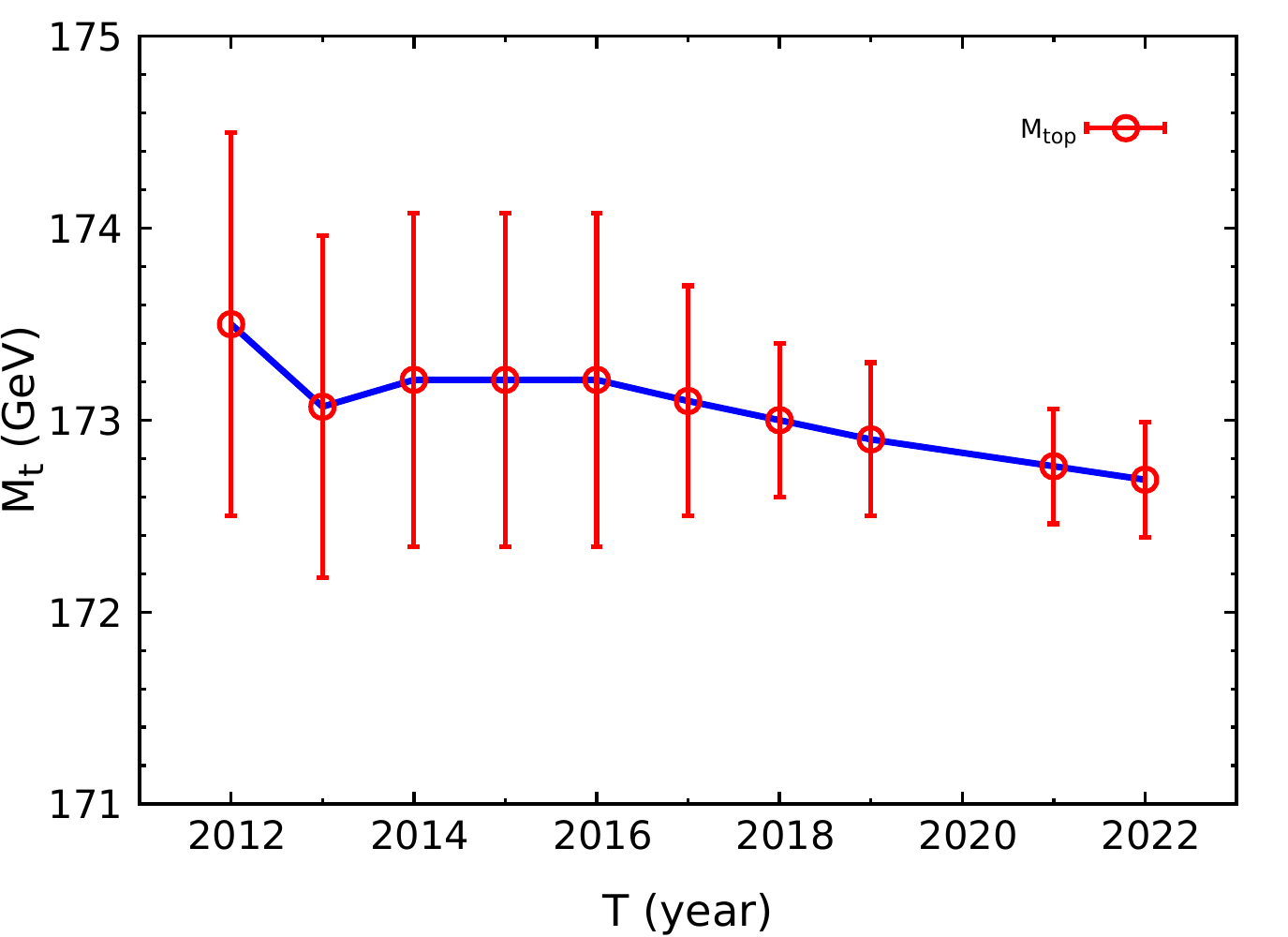}
    \caption{History of $M_t$ (top quark pole mass).}
    \label{fig:M_t}
  \end{subfigure}
  \hfill
  \begin{subtable}{0.45\linewidth}
    \vspace{-2mm}
    \renewcommand{\arraystretch}{1.17}
    \resizebox{0.99\linewidth}{!}{
    \begin{tabular}{@{\qquad} l @{\qquad} l @{\qquad} l }
      \hline\hline
      source & error (\%) & memo 
      \\ \hline
      $\Vcb$          & 49.7             & Exclusive \\
      $\eta_{ct}$     & 20.7             & $c-t$ Box \\        
      $\bar\eta$      & 13.3             & AOF \\
      $\eta_{cc}$     & \phantom{0}8.7   & $c-c$ Box \\        
      $\xi_\text{LD}$ & \phantom{0}2.1   & RBC-UKQCD \\        
      $\bar\rho$      & \phantom{0}2.1   & AOF \\        
      $\BK$           & \phantom{0}1.7   & FLAG \\
      $\;\vdots$      & $\;\;\;\;\vdots$ & $\;\;\vdots$
      \\ \hline\hline
    \end{tabular}
    } 
    \vspace*{4mm}
    \caption{Error budget for $|\epsK|^\text{SM}$}
    \label{tab:err-bud}
  \end{subtable}
  \caption{(\subref{fig:M_t}) $M_t$ history (\subref{tab:err-bud})
    error budget. }
  \label{tab:M_t+err_bud}
\end{table}

In Table \ref{tab:M_t+err_bud}\;(\subref{fig:M_t}), we plot top pole
mass $M_t$ as a function of time.
Here we find that the average value drifts downward a little bit and
the error shrinks fast as time goes on, thanks to accumulation of high
statistics in the LHC experiments.
The data for 2020 is dropped out intentionally to reflect on the
absence of Lattice 2020 due to COVID-19.

\section{$W$ boson mass}
\begin{table}[t!]
  \begin{subfigure}{0.55\linewidth}
    \vspace{-3mm}
    \includegraphics[width=1.0\linewidth]{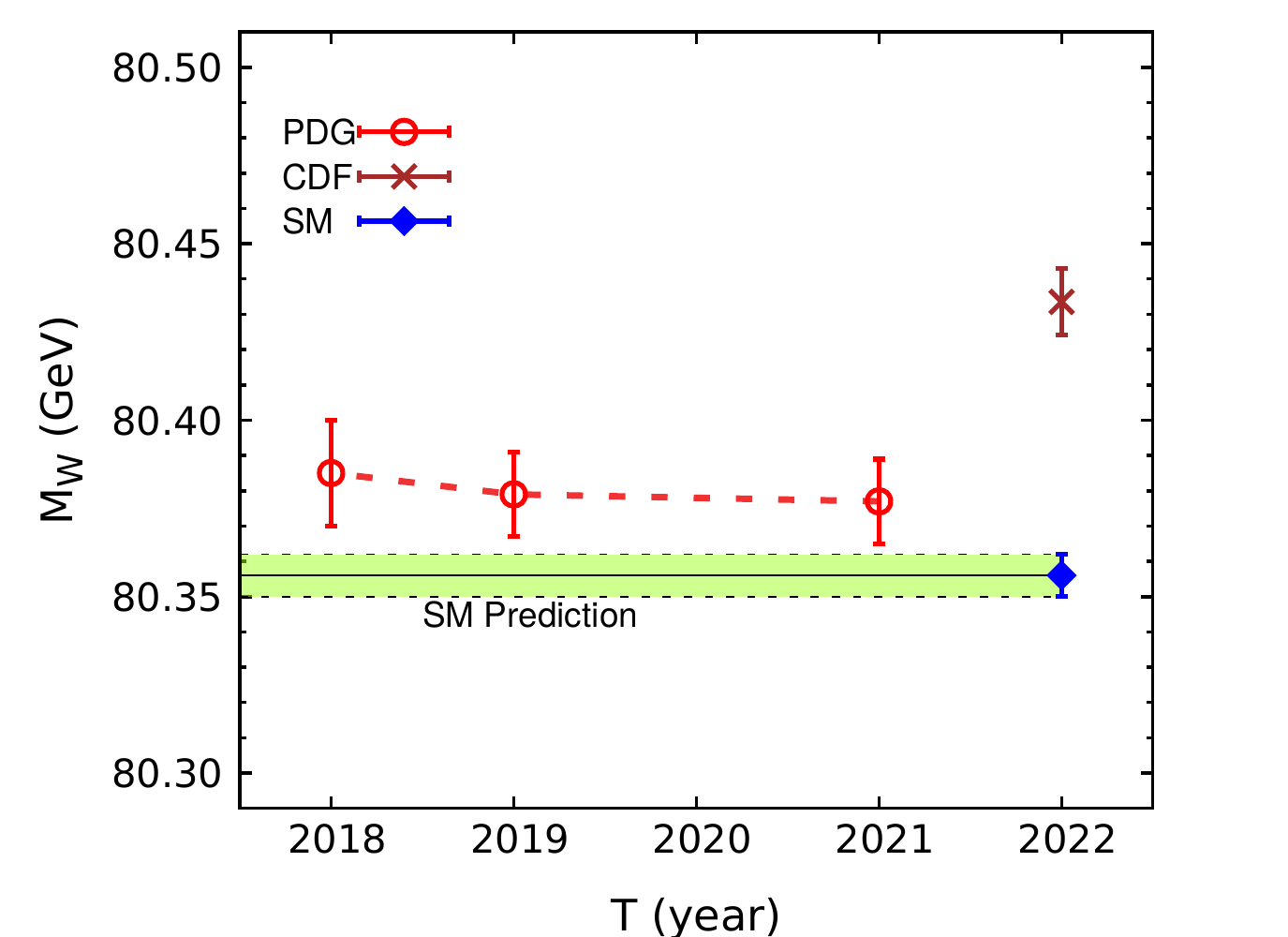}
    \caption{History of $M_W$ ($W$ boson mass).}
    \label{fig:m_W}
  \end{subfigure}
  \hfill
  \begin{subtable}{0.45\linewidth}
    \vspace{-2mm}
    \renewcommand{\arraystretch}{1.4}
    \resizebox{0.99\linewidth}{!}{
    \begin{tabular}{@{\qquad} l @{\qquad} l @{\qquad} l @{\qquad} }
      \hline\hline
      Source & $M_W$ (GeV) & Ref. 
      \\ \hline
      SM-2022  &  \red{80.356(6)}    & \cite{ Workman:2022ynf}  \\
      CDF-2022 & \blue{80.4335(94)}  & \cite{ CDF:2022hxs}  \\
      PDG-2021 &       80.377(12)    & \cite{ Zyla:2020zbs}  \\
      PDG-2019 &       80.379(12)    & \cite{ Tanabashi:2018oca}  \\
      PDG-2018 &       80.385(15)    & \cite{ Patrignani:2016xqp}
      \\ \hline\hline
    \end{tabular}
    } 
    \vspace*{4mm}
    \caption{Table of $M_W$ }
    \label{tab:m_W}
  \end{subtable}
  \caption{(\subref{fig:m_W}) $M_W$ history (\subref{tab:m_W})
    table of $M_W$. }
  \label{tab:m_W+fig}
\end{table}

In Fig.~\ref{tab:m_W+fig}\;(\subref{fig:m_W}), we plot $M_W$ ($W$
boson mass) as a function of time. The corresponding results for $M_W$
are summarized in Table \ref{tab:m_W+fig}\;(\subref{tab:m_W}).
In Fig.~\ref{tab:m_W+fig}\;(\subref{fig:m_W}), the light-green band
represents the standard model (SM) prediction, the red circles
represents the PDG results, and the brown cross represents the
CDF-2022 result.
The upside is that the CDF-2022 result is the most precise and latest
experimental result for $M_W$.
The downside, however, is that it has a $6.9\sigma$ tension from that
of SM-2022 (the standard model prediction).
Here, we use the SM-2022 result for $M_W$ to evaluate $\epsK$.

\section{Results for $\epsK$}
\begin{figure}[t!]
  \begin{subfigure}{0.47\linewidth}
    \vspace*{-5mm}
    \includegraphics[width=1.0\linewidth]
       {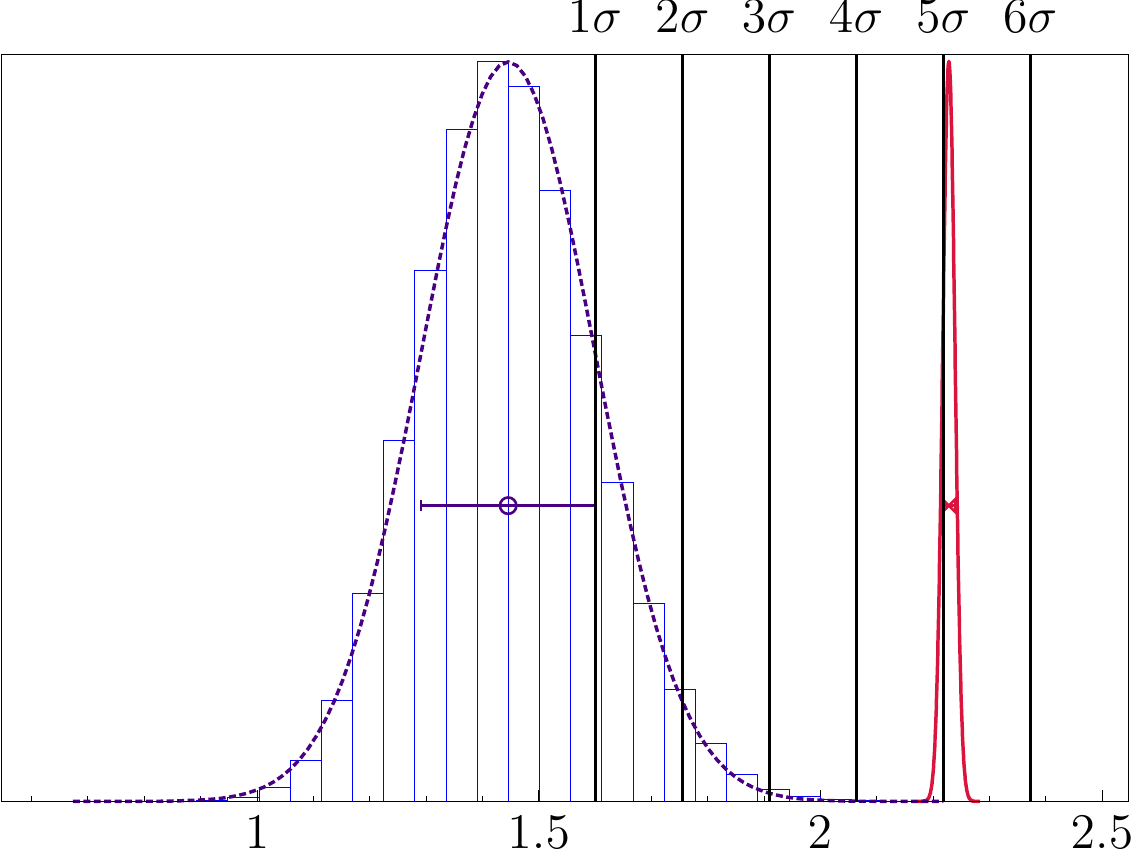}
    \caption{Exclusive $\Vcb$ (FNAL/MILC 2021, BGL)}
    \label{fig:epsK-ex:rbc}
  \end{subfigure}
  \hfill
  \begin{subfigure}{0.47\linewidth}
    \vspace*{-5mm}
    \includegraphics[width=1.0\linewidth]
                    {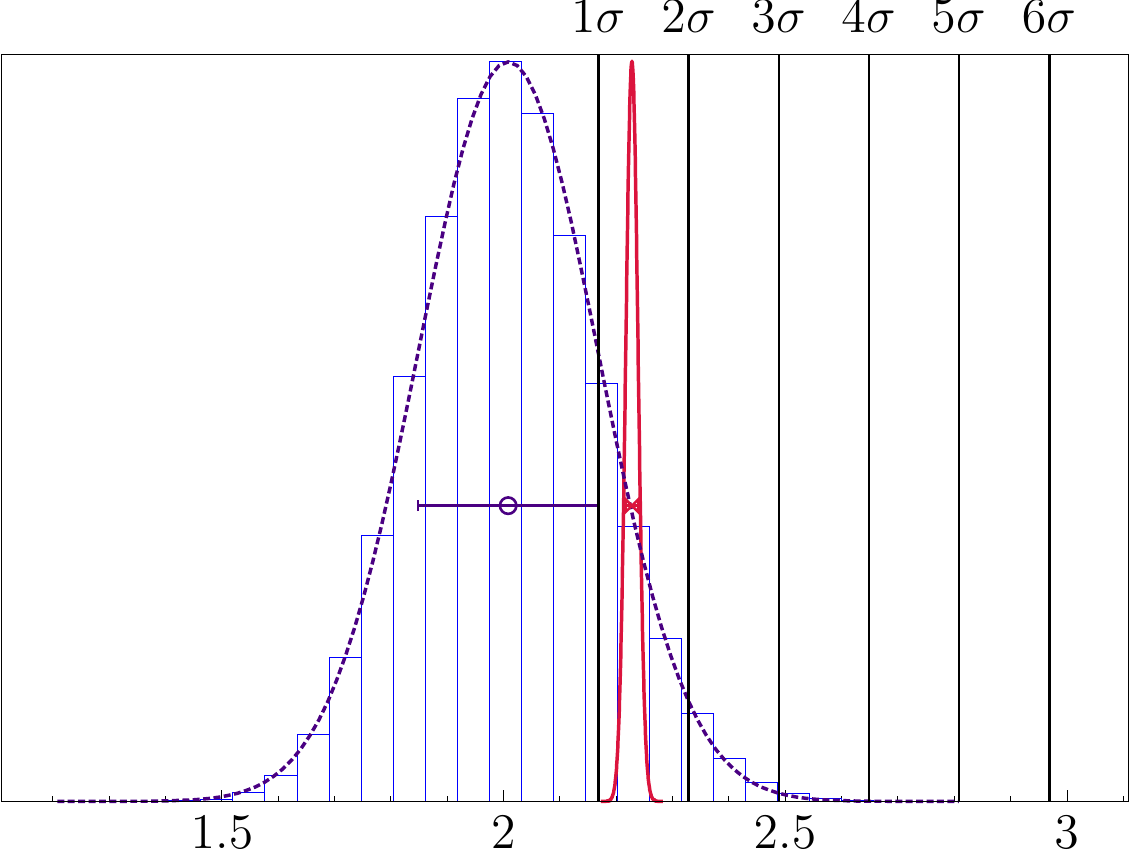}
    \caption{Inclusive $\Vcb$ (HFLAV 2021, 1S scheme)}
    \label{fig:epsK-in:rbc}
  \end{subfigure}
  \caption{$|\epsK|$ with (\subref{fig:epsK-ex:rbc}) exclusive $\Vcb$
    (left) and (\subref{fig:epsK-in:rbc}) inclusive $\Vcb$ (right) in
    units of $1.0\times 10^{-3}$. }
  \label{fig:epsK:cmp:rbc}
\end{figure}

In Fig.~\ref{fig:epsK:cmp:rbc}, we show results for $|\epsK|$
evaluated directly from the standard model (SM) with lattice QCD
inputs given in the previous sections.
In Fig.~\ref{fig:epsK:cmp:rbc}\;(\subref{fig:epsK-ex:rbc}), the blue
curve represents the theoretical evaluation of $|\epsK|$ obtained
using the FLAG-2021 results for $\BK$, AOF for Wolfenstein parameters,
the [FNAL/MILC 2022, BGL] results for exclusive $\Vcb$, results for
$\xi_0$ with the indirect method, and the RBC-UKQCD estimate for
$\xi_\text{LD}$.
The red curve in Fig.~\ref{fig:epsK:cmp:rbc} represents the experimental
results for $|\epsK|$.
In Fig.~\ref{fig:epsK:cmp:rbc}\;(\subref{fig:epsK-in:rbc}), the blue
curve represents the same as in
Fig.~\ref{fig:epsK:cmp:rbc}\;(\subref{fig:epsK-ex:rbc}) except for
using the 1S scheme results for the inclusive $\Vcb$.

Our results for $|\epsK|^\text{SM}$ and $\Delta\epsK$ are summarized
in Table \ref{tab:epsK}.
Here, the superscript ${}^\text{SM}$ represents the theoretical
expectation value of $|\epsK|$ obtained directly from the SM.
The superscript ${}^\text{Exp}$ represents the experimental value
of $|\epsK| = 2.228(11) \times 10^{-3}$.
Results in Table \ref{tab:epsK}\;(\subref{tab:epsK:rbc}) are obtained
using the RBC-UKQCD estimate for $\xi_\text{LD}$, and those in
Table \ref{tab:epsK}\;(\subref{tab:epsK:bgi}) are obtained using
the BGI estimate for $\xi_\text{LD}$.
In Table \ref{tab:epsK}\;(\subref{tab:epsK:rbc}), we find that the
theoretical expectation values of $|\epsK|^\text{SM}$ with lattice QCD
inputs (with exclusive $\Vcb$) has $5.12\sigma \sim 3.93\sigma$ tension
with the experimental value of $|\epsK|^\text{Exp}$, while there is no
tension with inclusive $\Vcb$ (obtained using heavy quark expansion
and QCD sum rules).
We also find that the tension with inclusive $\Vcb$ is small but keeps
increasing with respect to time.

\begin{table}[b!]
%
  \begin{subtable}{1.0\linewidth}
    \center
    \renewcommand{\arraystretch}{1.2}
    \resizebox{0.85\linewidth}{!}{
      \begin{tabular}{@{\qquad} l @{\qquad} l @{\qquad} l @{\qquad} l @{\qquad} l @{\qquad} }
        \hline\hline
        $\Vcb$    & method   & reference & $|\epsK|^\text{SM}$ & $\Delta\epsK$
        \\ \hline
        exclusive & BGL      & BELLE 2021 & $1.518 \pm 0.180$  & $3.93\sigma$
        \\
        exclusive & CLN      & BELLE 2021 & $1.532 \pm 0.171$  & $4.07\sigma$
        \\ \hline
        exclusive & BGL      & BABAR 2019 & $1.441 \pm 0.166$  & $4.72\sigma$
        \\ 
        exclusive & CLN      & BABAR 2019 & $1.446 \pm 0.161$  & $4.86\sigma$
        \\ \hline
        exclusive & BGL      & FNAL/MILC 2021 & $1.446 \pm 0.154$ & $5.05\sigma$
        \\
        exclusive & CLN      & HFLAV 2021 & $1.566 \pm 0.142$  & $4.63\sigma$
        \\ \hline\hline
        inclusive & kinetic  & Gambino 2021  & $2.041 \pm 0.168$ & $1.12\sigma$
        \\
        inclusive & 1S       & HFLAV 2021 & $2.008 \pm 0.160$ & $1.37\sigma$
        \\ \hline\hline
      \end{tabular}
    } 
    \caption{RBC-UKQCD estimate for $\xi_\text{LD}$}
    \label{tab:epsK:rbc}
  \end{subtable} 
  \begin{subtable}{1.0\linewidth}
    \vspace*{3mm}
    \center
    \renewcommand{\arraystretch}{1.2}
    \resizebox{0.85\linewidth}{!}{
      \begin{tabular}{@{\qquad} l @{\qquad} l @{\qquad} l @{\qquad} l @{\qquad} l @{\qquad} }
        \hline\hline
        $\Vcb$    & method   & reference  & $|\epsK|^\text{SM}$ & $\Delta\epsK$
        \\ \hline
        exclusive & BGL & FNAL/MILC 2021 & $1.494 \pm 0.157$ & $4.66\sigma$
        \\
        exclusive & CLN & HFLAV 2021     & $1.614 \pm 0.145$ & $4.22\sigma$
        \\ \hline\hline
      \end{tabular}
    } 
    \caption{BGI estimate for $\xi_\text{LD}$}
    \label{tab:epsK:bgi}
  \end{subtable} 
  \caption{ $|\epsK|$ in units of $1.0\times 10^{-3}$, and
    $\Delta\epsK = |\epsK|^\text{Exp} - |\epsK|^\text{SM}$.}
  \label{tab:epsK}
\end{table}

\begin{figure}[t!]
  \begin{subfigure}{0.501\linewidth}
    \includegraphics[width=\linewidth]
                    {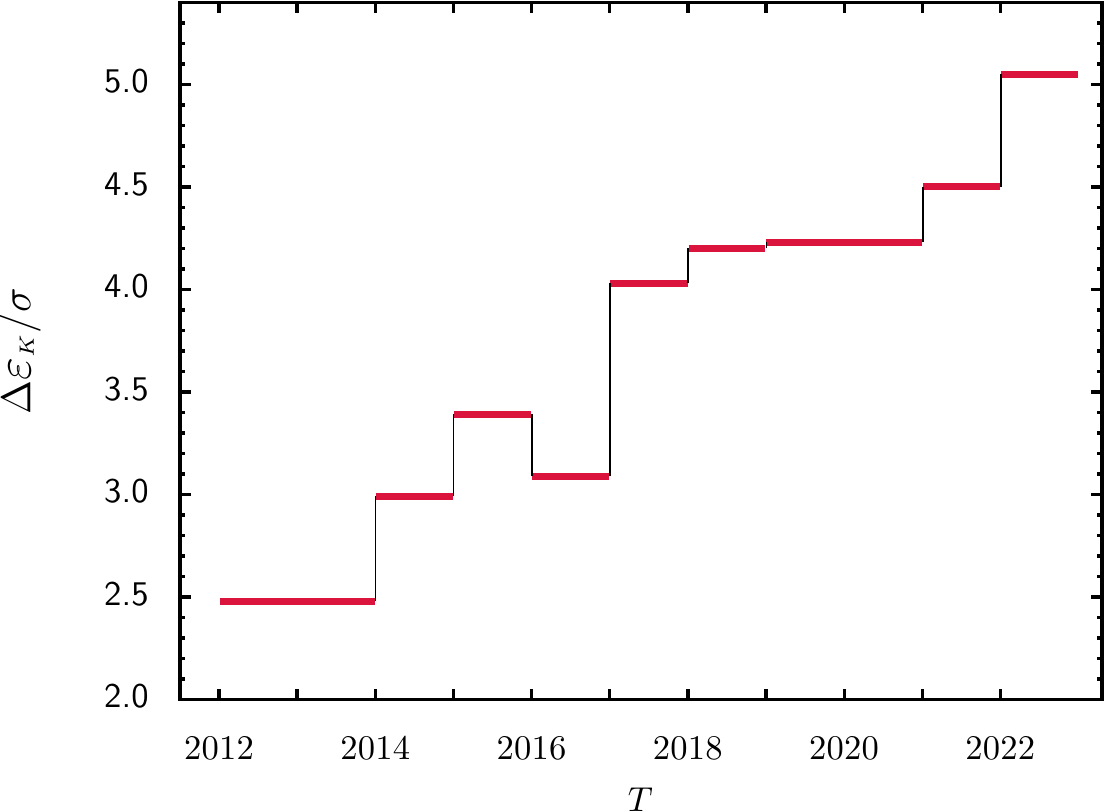}
    \caption{Time evolution of $\Delta \epsK/\sigma$}
    \label{fig:depsK:rbc:his}
  \end{subfigure}
  \hfill
  \begin{subfigure}{0.479\linewidth}
    \includegraphics[width=\linewidth]
                    {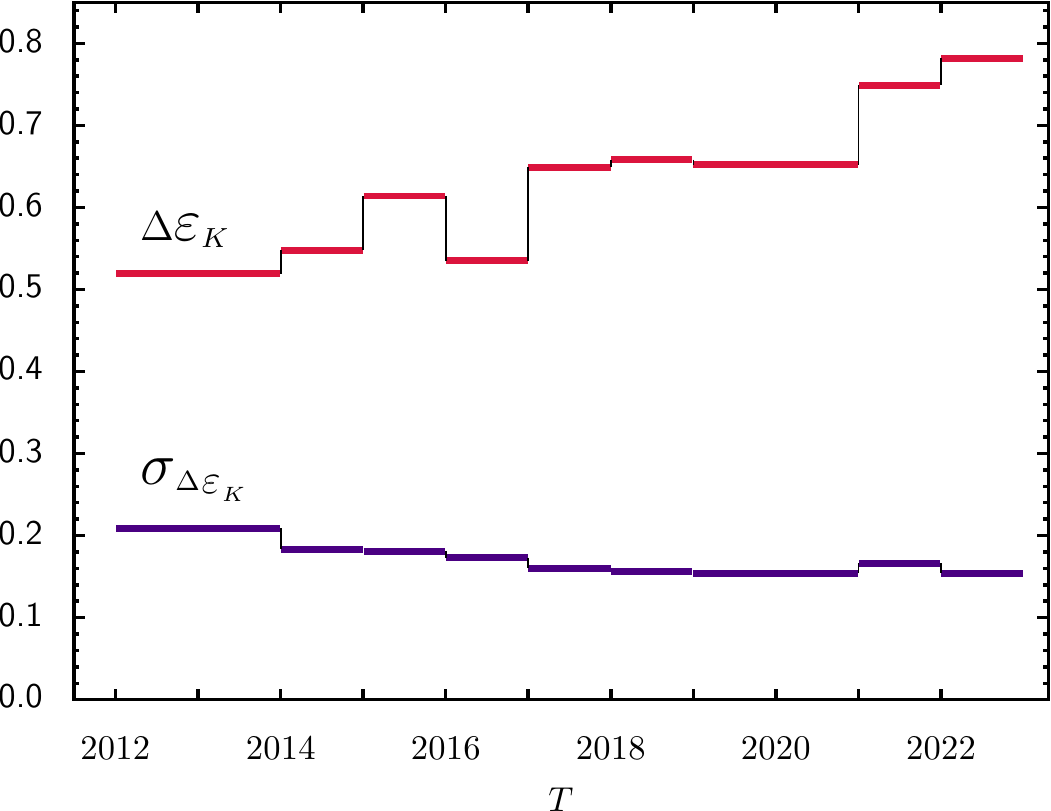}
    \caption{Time evolution of the average and error of $\Delta\epsK$}
    \label{fig:depsK+sigma:rbc:his}
  \end{subfigure}
  \caption{ Time history of (\subref{fig:depsK:rbc:his})
    $\Delta\epsK/\sigma$, and (\subref{fig:depsK+sigma:rbc:his})
    $\Delta\epsK$ and $\sigma_{\Delta\epsK}$. }
  \label{fig:depsK:sum:rbc:his}
\end{figure}

In Fig.~\ref{fig:depsK:sum:rbc:his}\;(\subref{fig:depsK:rbc:his}), we
show the time evolution of $\Delta\epsK$ starting from 2012 till 2022.
In 2012, $\Delta\epsK$ was $2.5\sigma$, but now it is $5.05\sigma$ with
exclusive $\Vcb$ (FNAL/MILC-2022, BGL).\footnote{Here, we use the
results for exclusive $\Vcb$ from FNAL/MILC-2022, since it contains
the most comprehensive analysis on the $\BtoDst$ decays on both zero
recoil and non-zero recoil data points, while it covers both BELL and
BABAR experimental results.}
In Fig.~\ref{fig:depsK:sum:rbc:his}\;(\subref{fig:depsK+sigma:rbc:his}),
we show the time evolution of the average $\Delta\epsK$ and the error
$\sigma_{\Delta\epsK}$ during the period of 2012--2022.

At present, we find that the largest error ($\approx 50\%$) in
$|\epsK|^\text{SM}$ comes from $\Vcb$.\footnote{Refer to Table
\ref{tab:M_t+err_bud} (\subref{tab:err-bud}) for more details.}
Hence, it is essential to reduce the errors in $\Vcb$ as much as
possible.
To achieve this goal, there is an on-going project to extract
exclusive $\Vcb$ using the Oktay-Kronfeld (OK) action for the heavy
quarks to calculate the form factors for $\BtoDstp$ decays \cite{
  Bhattacharya:2021peq, Park:2020vso, Bhattacharya:2020xyb,
  Bhattacharya:2018ibo, Bailey:2017xjk, Bailey:2017zgt,
  Bailey:2020uon}.

A large portion of interesting results for $|\epsK|^\text{SM}$ and
$\Delta\epsK$ could not be presented in Table \ref{tab:epsK} and in
Fig.~\ref{fig:depsK:sum:rbc:his} due to lack of space: for example,
results for $|\epsK|^\text{SM}$ obtained using exclusive $\Vcb$ (FLAG
2021), results for $|\epsK|^\text{SM}$ obtained using $\xi_0$
determined by the direct method, and so on.
We plan to report them collectively in Ref.~\cite{ wlee:2022epsK}.

%
%

\acknowledgments
We thank Jon Bailey, Yong-Chull Jang, Stephen Sharpe, and Rajan Gupta
for helpful discussion.
We thank Guido Martinelli for providing us the  most updated results of
the UTfit Collaboration in time.
The research of W.~Lee is supported by the Mid-Career Research
Program (Grant No.~NRF-2019R1A2C2085685) of the NRF grant funded by
the Korean government (MSIT).
W.~Lee would like to acknowledge the support from the KISTI
supercomputing center through the strategic support program for the
supercomputing application research (No.~KSC-2018-CHA-0043,
KSC-2020-CHA-0001).
Computations were carried out in part on the DAVID cluster at Seoul
National University.

\bibliography{refs}


\end{document}